\documentclass[preprint,12pt]{elsarticle}

\usepackage{graphicx}

\usepackage{amssymb}

\journal{JQSRT}

\begin{document}
\graphicspath{}

\begin{frontmatter}



\title{Evaluating the summer night sky brightness at a research field site on Lake Stechlin in northeastern Germany}


\author[label1,label2]{Andreas Jechow}
\author[label1]{Franz H{\"o}lker}
\author[label3]{Zolt{\'a}n Koll{\'a}th}
\author[label4,label5]{Mark O. Gessner}
\author[label1,label2]{Christopher C. M. Kyba}

\address[label1]{Ecohydrology, Leibniz Institute of Freshwater Ecology and Inland Fisheries (IGB), M{\"u}ggelseedamm 310, 12587 Berlin, Germany}
\address[label2]{Remote Sensing, GFZ German Research Centre for Geosciences, Telegrafenberg, 14473 Potsdam, Germany}
\address[label3]{University of West Hungary, Savaria Campus, Szombathely, Hungary}
\address[label4]{Experimental Limnology, Leibniz Institute of Freshwater Ecology and Inland Fisheries (IGB), Alte Fischerh{\"u}tte 2, 16775 Stechlin, Germany}
\address[label5]{Department of Ecology, Berlin Institute of Technology (TU Berlin), Ernst-Reuter-Platz 1, 10587 Berlin, Germany}

\begin{abstract}
We report on luminance measurements of the summer night sky at a field site on a freshwater lake in northeastern Germany (Lake Stechlin) to evaluate the amount of artificial skyglow from nearby and distant towns in the context of a planned study on light pollution. The site is located about 70 km north of Berlin in a rural area possibly belonging to one of the darkest regions in Germany. Continuous monitoring of the zenith sky luminance between June and September 2015 was conducted utilizing a Sky Quality Meter. With this device, typical values for clear nights in the range of 21.5-21.7 mag$_{SQM}/$arcsec$^2$ were measured, which is on the order of the natural sky brightness during starry nights. On overcast nights, values down to 22.84 mag$_{SQM}/$arcsec$^2$ were obtained, which is about one third as bright as on clear nights. The luminance measured on clear nights as well as the darkening with the presence of clouds indicate that there is very little influence of artificial skyglow on the zenith sky brightness at this location. Furthermore, fish-eye lens sky imaging luminance photometry was performed with a digital single-lens reflex camera on a clear night in the absence of moonlight. The photographs unravel several distant towns as possible sources of light pollution on the horizon. However, the low level of artificial skyglow makes the field site at Lake Stechlin an excellent location to study the effects of skyglow on a lake ecosystem in a controlled fashion. 
\end{abstract}


\end{frontmatter}


\section{Introduction}
Since the invention of electric lighting in the 19th century, artificial light at night (ALAN) has had a major impact on both human society and the environment. ALAN propagates into nocturnal landscapes and has hence been recognized as one type of environmental pollution, light pollution (LP). The fact that ALAN can be seen from space should be well known since its observation during the first crewed US American orbital flight more than 50 years ago \cite{Biggs:2012}. However, the public is much less aware of LP than of other types of environmental pollution. Although ALAN has globally increased by 3-6$\%$ per year \cite{Hoelker:2010_b, book:Narisada} over the last decades, the effects on the environment and on human well-being are still poorly understood \cite{Hoelker:2010_b}.

Early work on LP concentrated mostly on the decrease of contrast of the night sky and the resulting lack of the ability to observe stars during the night (see \cite{book:Narisada} for an overview). Later it was found that LP has strong effects on flora, fauna \cite{book:rich_longcore} and human well-being \cite{Stevens:2015}. Recently, concerns were raised that LP might have an impact on whole ecosystems and biodiversity, and several investigations in that context have been performed \cite{Hoelker:2010_a, Gaston:2015_ptb}. While most work has focused on terrestrial animals, some recent work investigated the influence of ALAN on aquatic life \cite{Perkin:2011}. This includes studies on the diel vertical migration of zooplankton in lakes \cite{Moore:2000}, the alteration of microbial communities \cite{Hoelker:2015} and photophysiology of freshwater cyanobacteria \cite{Poulin:2014}. Since ALAN is increasingly recognized as an environmental stressor, several studies to quantify the amount of ALAN leaving the Earth's surface have been performed. A combination of satellite images and theoretical modeling resulted in the World Atlas of night sky brightness \cite{Cinzano:2001_wa}.

Artificial skyglow is one type of LP that is mainly observed in densely populated areas where ALAN is scattered by atmospheric molecules or aerosols and returned to Earth \cite{Aube:2015}. Skyglow can be drastically amplified by several orders of magnitude due to the presence of clouds as shown by empirical \cite{Kyba:2011_sqm, Puschnig:2014_Vienna} and theoretical \cite{Kocifaj:2014_cloud_theory} work. This amplification depends not only on the amount of upwelling radiation caused by ALAN but also on its spectral distribution, the topography of the landscape, the vegetation cover and especially the weather condition \cite{Kyba:2012_mssqm}.

Despite its possible impact on ecosystems \cite{Kyba:2013_LE}, it is not straightforward to estimate the amount of skyglow at a specific location from satellite images alone, as these are only usable in clear conditions. Furthermore, the seminal World Atlas of night sky brightness \cite{Cinzano:2001_wa} is based on almost 20 year old satellite data; and unfortunately no such updated global data base is available, yet. ALAN however, is rapidly changing especially very recently due to the availability of cheap and efficient solid state lighting. As a result, prediction of the current exposure to LP at a field site, particularly for overcast conditions, is difficult.

This lack of reliable data, together with the development of cheap handheld sky brightness meters \cite{Cinzano_sqm:2005} and novel measurement schemes based on imaging technology \cite{Duriscoe:2007, Kollath:2010} have triggered several local \cite{Biggs:2012, Pun:2014} and global \cite{Kyba:2015_isqm} ground-based studies of the night sky brightness (NSB). In this work, we measured the NSB at Lake Stechlin over a period of four months in summer and early autumn to evaluate the amount of artificial skyglow at an experimental field site. We employed continuous four-month measurements with a sky quality meter (SQM) as well as imaging with a digital single lens reflex (DSLR) camera. The motivation of our work is to characterize skyglow at a field site where an experiment is planned to assess the impact of LP on a lake ecosystem. For this experiment it has to be verified that the field site is located in one of the darkest regions of Germany \cite{Cinzano:2001_wa}, which makes it possible to use unmanipulated experimental units as controls.

\section{The field site}
\begin{figure}
\centering
(a) \includegraphics[width=13cm]{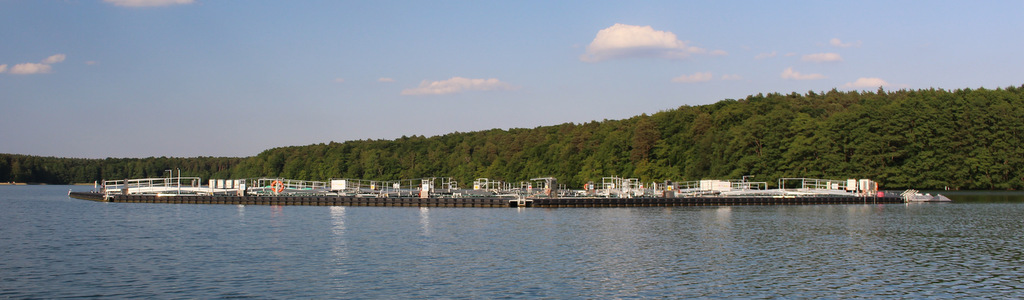}
(b) \includegraphics[width=13cm]{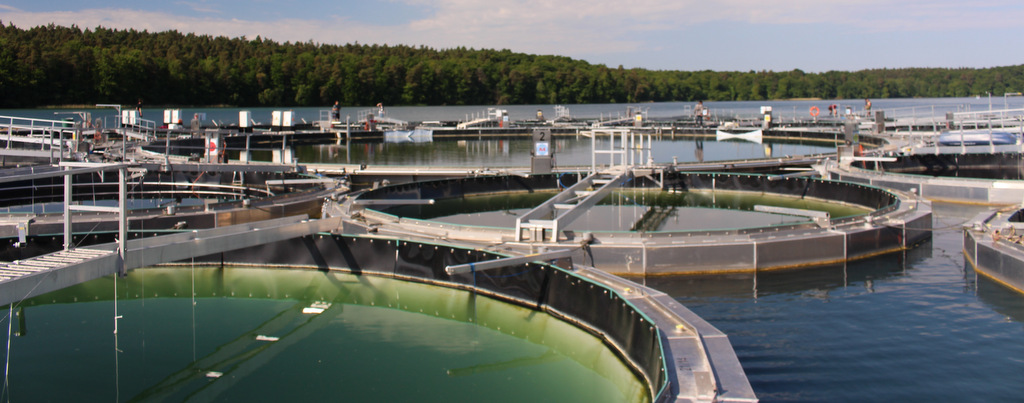}
\caption{(a) Distant and (b) close-up view of the experimental facility (LakeLab) in Lake Stechlin, Germany, where night sky brightness measurements were conducted. (Pictures by A. Jechow).}
\label{lakelab}
\end{figure}

Measurements were made at the LakeLab (Fig. \ref{lakelab}; www.lake-lab.de), which is an experimental mesocosm facility installed in Lake Stechlin, a clearwater lake with a surface area of 4.25 km² and a maximum depth of 69.5 m. The lake is situated about 70 km north of the city of Berlin in a forested area. The nearest village is Neuglobsow (53$^{\circ}$8$^{\prime}$48$^{\prime\prime}$N, 13$^{\circ}$2$^{\prime}$49$^{\prime\prime}$E) with less than 300 inhabitants. The lake is part of a nature reserve belonging to one of the darkest areas of Germany (according to \cite{Cinzano:2001_wa}).

The experimental facility is situated in the south-western bay of the lake (53$^{\circ}$8$^{\prime}$36.5$^{\prime\prime}$N, 13$^{\circ}$1$^{\prime}$42.7$^{\prime\prime}$E). The floating construction consists of one large enclosure  with 30 m diameter and 24 smaller enclosures that are 9 m in diameter (Fig. \ref{lakelab}). Each enclosure (mesocosm) is isolated from the lake by plastic curtains that reach down into the lake sediment at a depth of about 20 m. The entire structure is anchored to the lake bottom to prevent drift.

The purpose of the LakeLab is to assess impacts of global environmental change on ecological processes and biodiversity in lake ecosystems. The relatively large size of the experimental units (1270 m$^3$ each) and their large number (24) provide a rare opportunity to test for effects under controlled and yet realistic conditions by capturing most of the natural complexity of lake ecosystems, and at the same time ensuring adequate replication when only one or two factors are experimentally varied while all others are kept constant. Multiple environmental stressors can be tested including LP.  Thus, experiments in the LakeLab combine the advantages of field and laboratory experimental research. 

\section{Methods}
 \subsection{The Sky Quality Meter (SQM)}
Continuous night sky luminance measurements were performed between June and September 2015 with an SQM manufactured by Unihedron (Grimsby, Ontario, Canada). The SQM version with an integrated lens (SQM-Lxx) measures luminance for a patch of the sky with an opening angle of 20$^{\circ}$ \cite{Cinzano_sqm-l:2007}. The combination of a silicon photo diode (TSL237S) and a band-pass filter (HOYA CM-500) results in a spectral response roughly similar to that of the scotopic human vision \cite{Cinzano_sqm:2005}. The SQM provides the luminance in units of magnitude per square arc second (mag$_{SQM}$/arcsec$^2$), which is a logarithmic scale that decreases with increasing brightness. Because the spectral sensitivity does not match the Johnson V band, we followed the convention of Biggs and others \cite{Biggs:2012} to use mag$_{SQM}$/arcsec$^2$ (see also \cite{Cinzano_sqm:2005}). A value of 21.6 mag$_{SQM}$/arcsec$^2$, corresponds to the brightness (or darkness) of a typical historic moonless clear night with only natural contributions to the NSB \cite{Kollath:2010, Kyba:2015_isqm}. A decrease of 5 mag$_{SQM}$/arcsec$^2$ equals an increase in luminance by a factor of 100. This scale used in astronomy can be converted to so-called natural sky units (NSU) to provide a more intuitive comparison between different NSBs \cite{Kyba:2015_isqm}. One NSU indicates how much brighter or darker a sky is compared to a typical historic clear night sky (21.6 mag$_{SQM}$/arcsec$^2$; NSU=1). It is defined as 

\begin{equation}
NSU = 10^{0.4(21.6-x)},
\end{equation}
where x is the observed value in mag$_{SQM}$/arcsec$^2$.

SQMs are widely used, due to their low cost even in large-scale projects involving citizen scientists \cite{Kyba:2013_GaN}. A recent study collected and compared global data from SQMs to evaluate the variations in artificial skyglow at multiple locations experiencing different light pollution situations \cite{Kyba:2015_isqm}. During such long term measurements, the SQM proved to be robust and reliable. Due to an internal sensor and correction algorithm, it is almost insensitive to temperature variations in the range of -25$^{\circ}$C to 70$^{\circ}$C \cite{Schnitt:2013}.

\begin{figure}
\centering
\includegraphics[width=6cm]{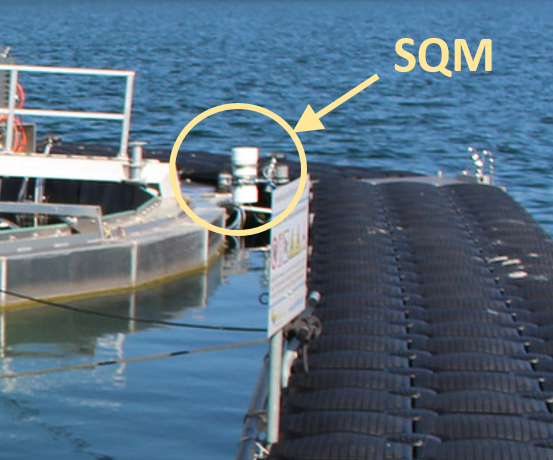}
\caption{Sky quality meter (SQM) mounted in a weather-proof housing on the floating platform at the experimental facility (LakeLab) in Lake Stechlin, Germany. (Picture by A. Jechow).}
\label{SQM}
\end{figure}
We used a battery-powered version of the SQM with an integrated data logger (SQM-LU-DL). The SQM was set to measure the sky luminance every 10 minutes. The device was integrated in a weather-proof housing (Unihedron) and was permanently mounted (Fig. \ref{SQM}) on the experimental facility in Lake Stechlin. Care was taken that the SQM pointed straight up to the sky.

\subsection{Fish-eye lens sky imaging with a digital single-lens reflex camera}
Modern DSLR cameras with the ability to save images in an unaltered raw format have recently been used to evaluate NSB \cite{Kollath:2010, Kocifaj:2015}. Calibration is necessary to translate the brightness values of the camera to physical luminance values. It is then possible to process false color images that show spatially resolved NSB in units of mag/arcsec$^2$ or NSU, respectively. If carefully calibrated, it is possible to reach 10 percent precision with such an imaging measurement system \cite{Kollath:2010}. 

We used a Canon EOS D6 camera and a Sigma EX DG circular fish-eye lens. The lens has a focal length of f=8 mm and an aperture of F3.5. The camera features a 20.2 Megapixel full frame (36 mm x 24 mm) CMOS sensor and an integrated GPS tracker. The camera was cross-calibrated with a thoroughly calibrated camera during an earlier measurement campaign \cite{Kyba:2015_lonne}. Pictures were obtained in full format (5472 pixel x 3648 pixel) with ISO1600 setting and an exposure time of 30 seconds and saved in raw format. As is standard in astrophotography, dark frames were taken with the same settings and subtracted from the data images. This reduces noise and provides the appropriate baseline. The images were processed using a homebuilt software and then saved in JPEG format with a quadratic frame of 900 pixel x 900 pixel.

Usually DSLR cameras are pointed to the zenith to obtain so called ``all-sky images'' \cite{Kollath:2010, Kocifaj:2015}. However, aberrations and distortion effects are most prominent at the edges of the images. As we are mostly interested in the brightness at the horizon we pointed the camera not only to the zenith but also approximately towards the horizon to avoid such negative effects.

\subsection{Weather conditions}
To estimate the cloud coverage at the field site, weather conditions were obtained from the OGIMET free data base (www.ogimet.com). For the LakeLab at Lake Stechlin, data from the closest weather station (Neuruppin ID 10270) at about 20 km distance was used.

\section{Results}
 \subsection{Continuous zenith sky brightness measurements}
\begin{figure}[!t]
\centering
\includegraphics[width=10cm]{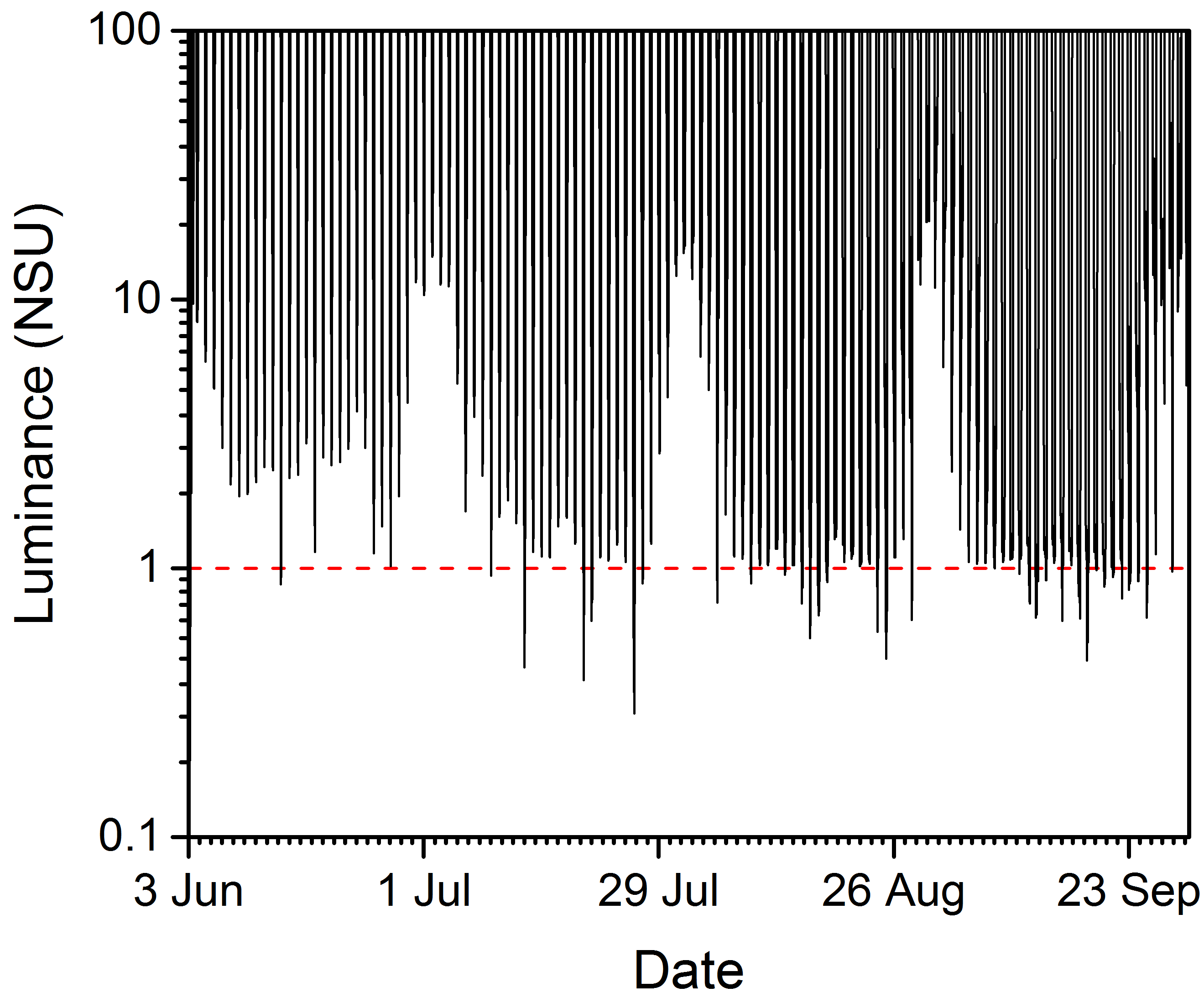}
\caption{Night sky brightness measurements performed with the Unihedron sky quality meter (SQM) from June 2015 to September 2015 with a temporal resolution of 10 minutes. The luminance is given in natural sky units (NSU) on a logarithmic scale. The red horizontal line represents the brightness of a historical clear night at 1 NSU (21.6 mag$_{SQM}/$arcsec$^2$). The influence of moon phases and weather conditions can be observed.}
\label{full}
\end{figure}

Figure \ref{full} shows the variation in sky luminance in NSU on a logarithmic scale as a function of time from the beginning of June until the end of September. NSU values were recorded automatically during twilight and the night, while the SQM saturates in bright conditions and therefore no data was recorded during daytime.

The diel change in sky brightness from nighttime to daytime is clearly visible in the plot. No filtering or averaging was used in the plot and the full data is shown. Each spike represents the lowest value obtained each night, while the gaps between the spikes are daylight periods.

It is also clear from Fig. \ref{full} that the moon phases change the NSB from bright periods at full moon (e.g. around 1 July 2015) to dark periods at new moon (e.g. around 15 July 2015). Lowest NSB values ranged from 0.32 NSU (22.84 mag$_{SQM}/$arcsec$^2$ on 26 July 2015 in cloudy conditions) to 10.4 NSU (1 July 2015 at full moon). The reading of about 10 NSU around full moon does not mean that moonlight is only 10 times brighter than starlight. Rather, the moon is a bright point source outside of the measured patch of the sky. The increased brightness readings are due to leaking of a small fraction of moonlight into the SQM optics or the secondary zenith sky brightening due to lunar twilight. For this reason, SQM measurements are usually analyzed only in the absence of moonlight.

\begin{figure}[!t]
\centering
\includegraphics[width=10cm]{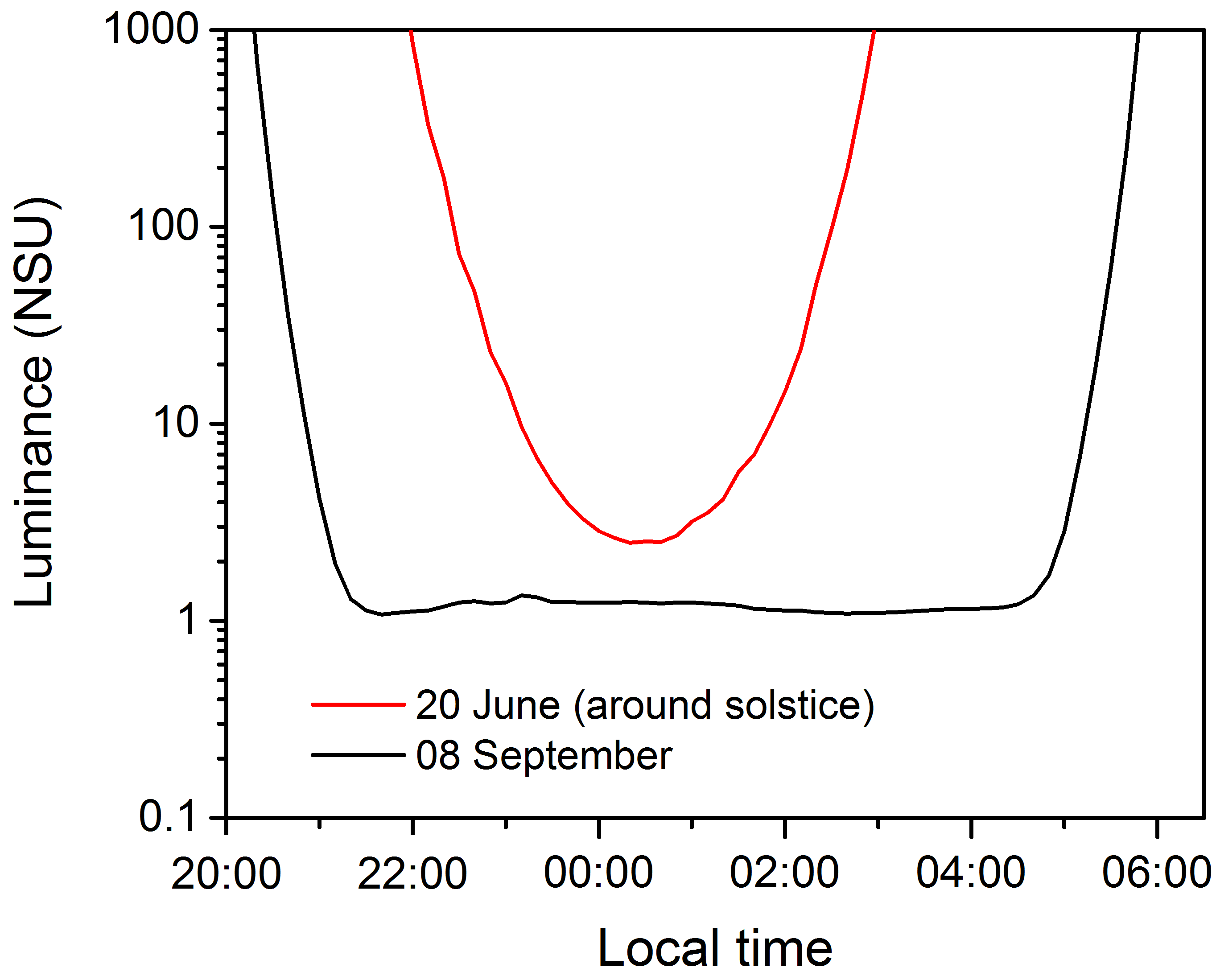}
\caption{Zenith night sky luminance at LakeLab measured with an SQM on a clear night around summer solstice (red line) and a clear night in September (black line). In fall, the night lasts about 7 hours at values around 1 NSU, while at summer solstice the night is not fully present as zenith brightness does not reach below 3 NSU.}
\label{nightlength}
\end{figure}

The minimum NSB depends not only on the moon phase, but also on the length of the day, which changes due to the axial tilt of the earth. For example astronomical twilight (defined as the period when the sun is between 12 $^{\circ}$ and 18 $^{\circ}$ below the horizon) does not end at the site between mid-May and end of July. This can be resolved in the data set when paying attention to the lowest NSB from the start of the observation until the end of July. The moonlight oscillation overlapped with a seasonal solar oscillation and the nights become brighter around solstice. In Fig. \ref{nightlength} two clear nights are shown around solstice (red) and later in September (black). The NSB at summer solstice does not reach below 3 NSU, and only a short period of about 3 hours below 10 NSU each night. In contrast, the clear night in September maintains values near 1 NSU from 21:00 to 05:00 H.

 \subsection{Influence of clouds on zenith sky brightness}
Night sky luminance can vary dramatically with the presence of clouds. In an environment without LP, clouds will darken the sky. However, in areas with LP clouds can make the sky considerably brighter: maximum values up to 2200 NSU have been reported \cite{Kyba:2015_isqm}. This effect is due to the broad and relatively high reflectance of clouds and is discussed in more detail in \cite{Lamphar:2015}.

\begin{figure}[!t]
\centering
\includegraphics[width=10cm]{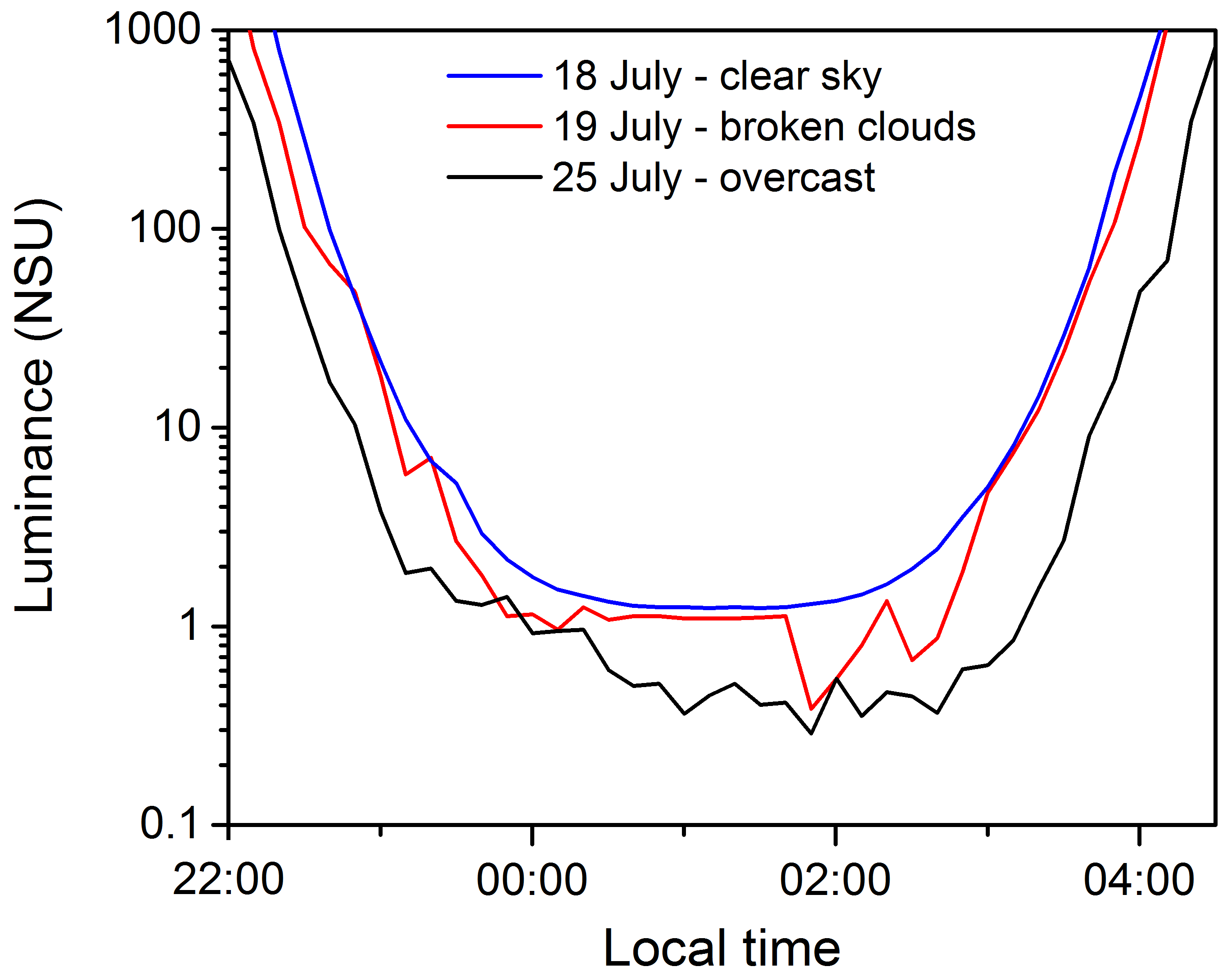}
\caption{Zenith night sky luminance at LakeLab measured with an SQM on a clear night (blue line), a night with broken clouds (red line) and an overcast night (black line). The presence of clouds make the sky darker.}
\label{clear_clouds}
\end{figure}

In our measurements the influence of clouds was clearly visible. Figure \ref{clear_clouds} shows two consecutive nights without (blue line: 18 July 2015 - moonset at 22:06 H) and with (red line: 19 July 2015 - moonset at 22:31 H) the presence of broken clouds as well as a third night which was completely overcast (black line: 25 July 2015 - moonset at 00:28). The consecutive nights have similar durations of about 4 hours with values below 10 NSU and a period of more than 2 hours with NSU values around 1. The influence of the lunar twilight for these two nights is hardly perceptible.

While the luminance values of the clear night smoothly approached the minimum of about 1.3 NSU, oscillations and jumps from darker to brighter sky luminance are apparent during the following night when broken clouds were present. This variation can be observed between 22:00 H and midnight but the most dramatic change occurred at around 02:00 H. Here the sky luminance dropped two times below values of 1 NSU before returning to the previous value of about 1.1-1.2 NSU. For the night with broken clouds (blue lin in plot) a value as low as 0.43 NSU (22.5 mag$_{SQM}$/arcsec$^2$) was observed, and for the completely overcast night (red line in plot) a value as low as 0.32 NSU (22.84 mag$_{SQM}$/arcsec$^2$) was observed. The slow decrease in brightness at the beginning of that night can presumably be attributed to moonlight and the following lunar twilight.

The data collected at the LakeLab indicates that clouds darken the night sky at this location, whereas the opposite would be observed with a significant amount of artificial skyglow caused by ALAN. This is illustrated by comparing the SQM data acquired at the LakeLab with that from other sites.
\begin{figure}[!b]
\centering
\includegraphics[width=6.75cm]{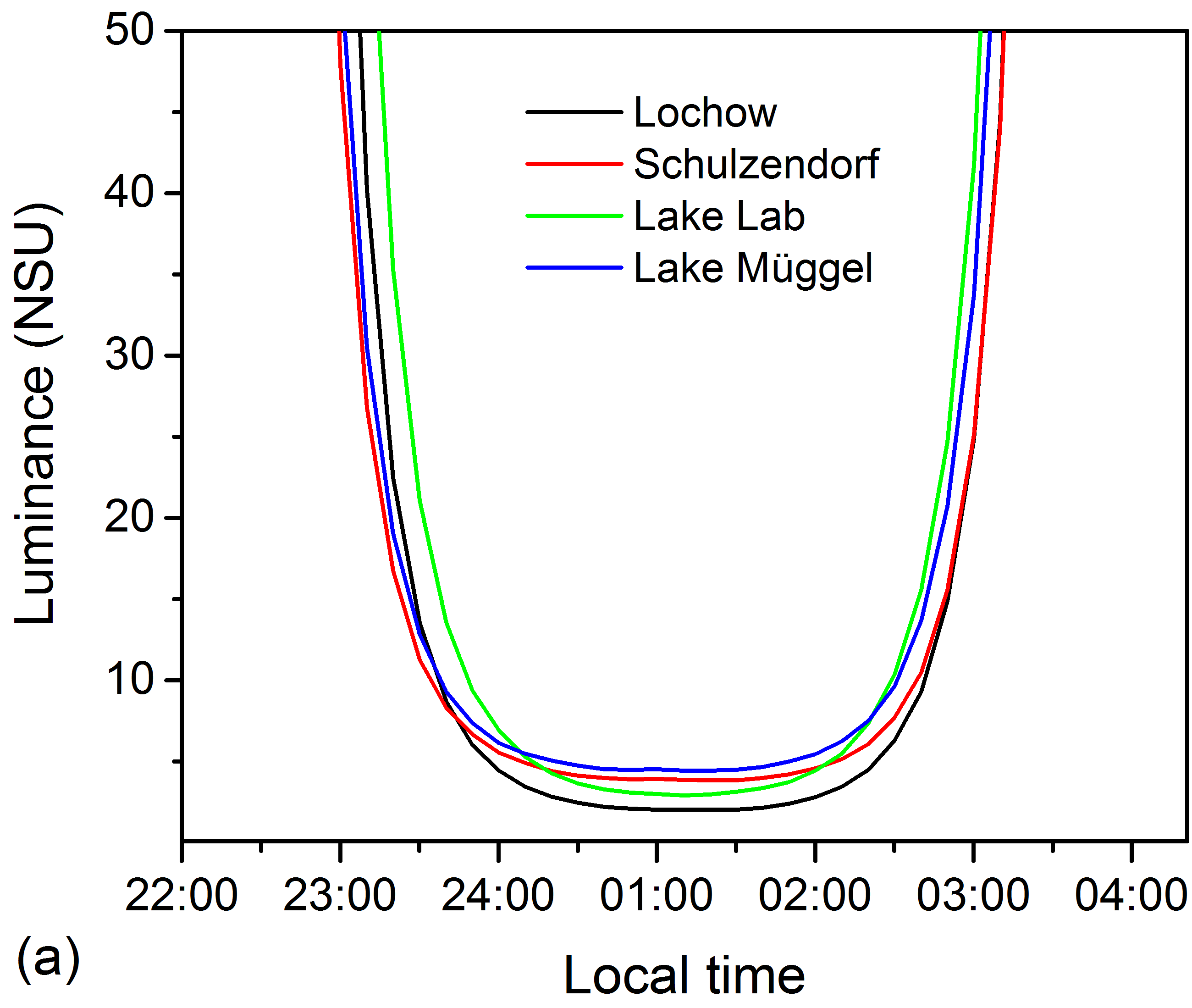}
\includegraphics[width=6.75cm]{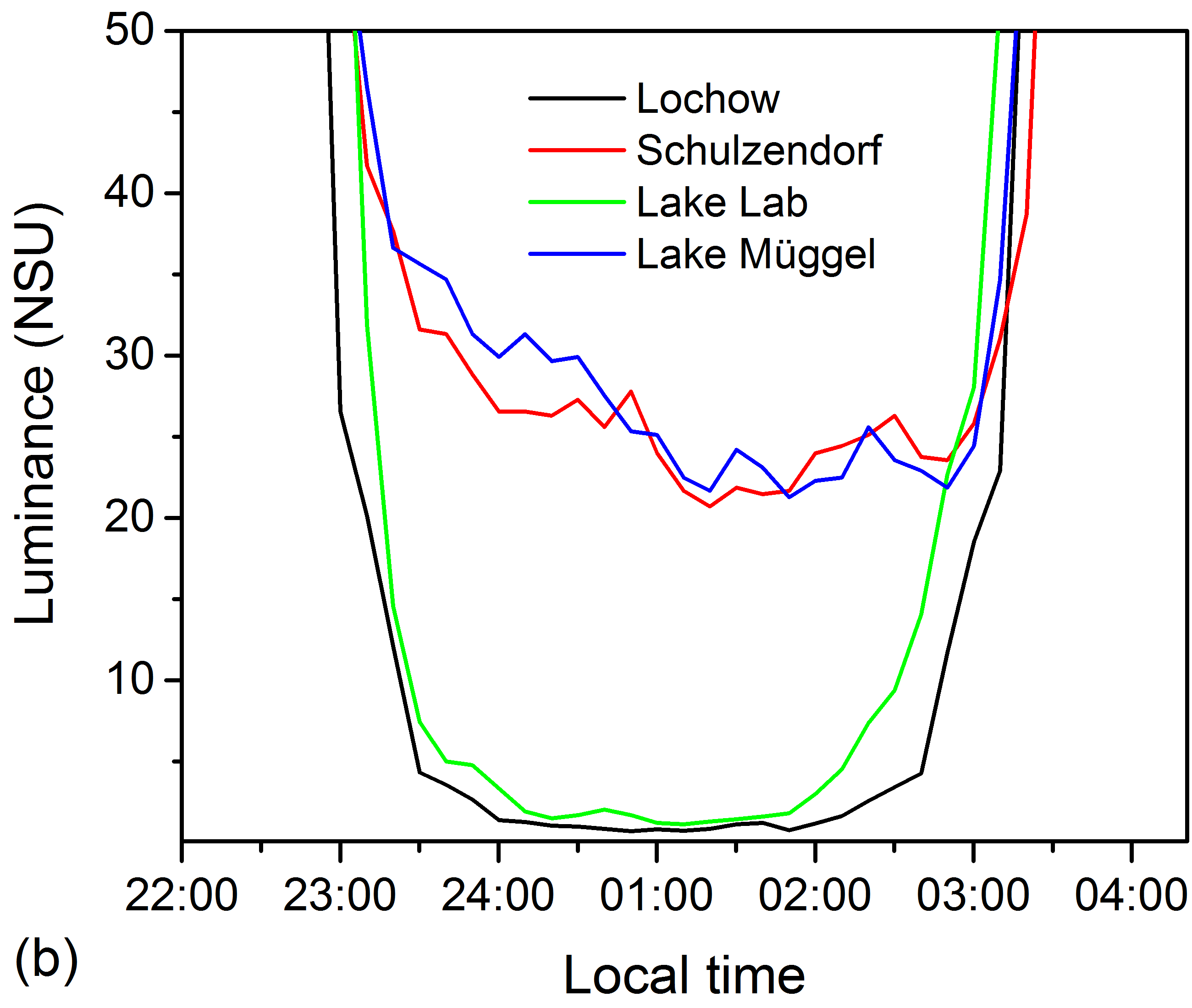}
\caption{Zenith night sky luminance on a clear night (a) and an overcast night (b) measured with SQMs at four different locations. In the presence of clouds a decrease in brightness is observed at the two rural locations Lochow (black line) and LakeLab (green line), while a drastic increase in brightness is observed at the two suburban locations IGB headquarter at Lake M{\"u}ggelsee (blue line) and Schulzendorf (red line). As a result, the NSB at the suburban sites is up to 50 times higher than at the rural sites, although the measurements were made in summer close to the solstice.}
\label{comp}
\end{figure}
Figure \ref{comp} shows the SQM data collected at four different locations for (a) a clear night (16 June2015) and (b) an overcast night (17 June 2015). One SQM was located at another experimental IGB field site \cite{Holzhauer2015}, which is situated in the International Dark-Sky Reserve Westhavelland about 70 km West of Berlin (52$^{\circ}$42$^{\prime}$N 12$^{\circ}$27$^{\prime}$E). The data from this SQM is labeled Lochow and shown by a black line in the plot. Two other SQMs are situated in a suburban setting at the outskirts of Berlin, one at the IGB headquarter at M{\"u}ggelsee (52$^{\circ}$26$^{\prime}$55$^{\prime\prime}$N 13$^{\circ}$38$^{\prime}$52$^{\prime\prime}$E) in the suburbs of Berlin and another one in Schulzendorf (52$^{\circ}$21$^{\prime}$36$^{\prime\prime}$N 13$^{\circ}$35$^{\prime}$24$^{\prime\prime}$E), a municipality close to the southeastern outskirts of Berlin (about 8,000 inhabitants).

The plots show NSB in a linear scale for two consecutive nights without moonlight. During the clear night (Fig. \ref{comp} a) the difference between the four locations is small but perceptible with values ranging from 1.1 NSU at Lochow (black) to 5 NSU at IGB headquarter at M{\"u}ggelsee (blue). However, these differences increase dramatically in the presence of clouds (Fig. \ref{comp} b). The two rural sites got darker (LakeLab and Lochow), whereas the two suburban sites (Schulzendorf and M{\"u}ggelsee) got much brighter than during the clear night. At Lochow, the night got as dark as 0.7 NSU while at Lake M{\"u}ggelsee at the outskirts of Berlin the LP was amplified by the clouds reaching values of more than 30 NSU before midnight and above 20 NSU after midnight. This decrease in brightness is characteristic as lamps are typically switched off as the night progresses \cite{Kyba:2011_sqm, Puschnig:2014_Vienna}. Both nights were very close to the summer solstice and the night in the Berlin area was very short (no astronomical twilight and nautical twilight from 23:45 H to about 2:30 H). Nevertheless, the zenith NSB was up to 50 times higher at the suburban than at the rural sites. For comparison, the NSB in the center of Berlin (Hackescher Markt) typically reaches values between 20 and 40 NSU on clear nights and between 250 and 750 NSU under overcast conditions \cite{Kyba:2015_isqm}.

\subsection{Fish-eye lens sky images}
Fish-eye lens sky images were taken during the night from 23 September 2015 to 24 September 2015 under clear weather conditions. On these dates, the moon rose at about 16:20 H local time and set at about 02:00 H at night. Astronomical twilight began at about 05:00 H in the morning on 24 September 2015. Therefore, a suitable time window for the NSB measurements was between 03:00 H and 05:00 H. NSB values obtained from the continuous four-month measurement with the SQM were in the range of 1.0-1.1 NSU between 02:00 H and 05:00 H, which was confirmed with a handheld SQM while taking the fish-eye lens images.

\begin{figure}
(a) \includegraphics[width=5.4cm]{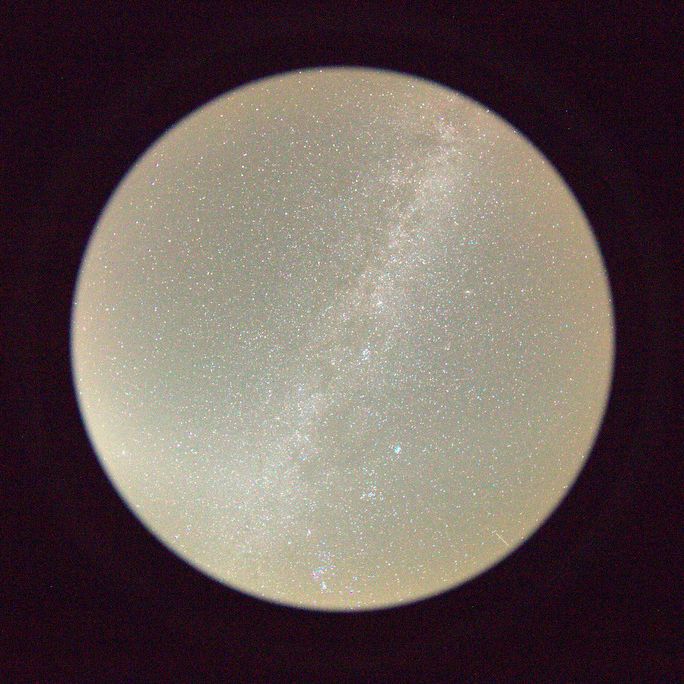}
(b) \includegraphics[width=5.4cm]{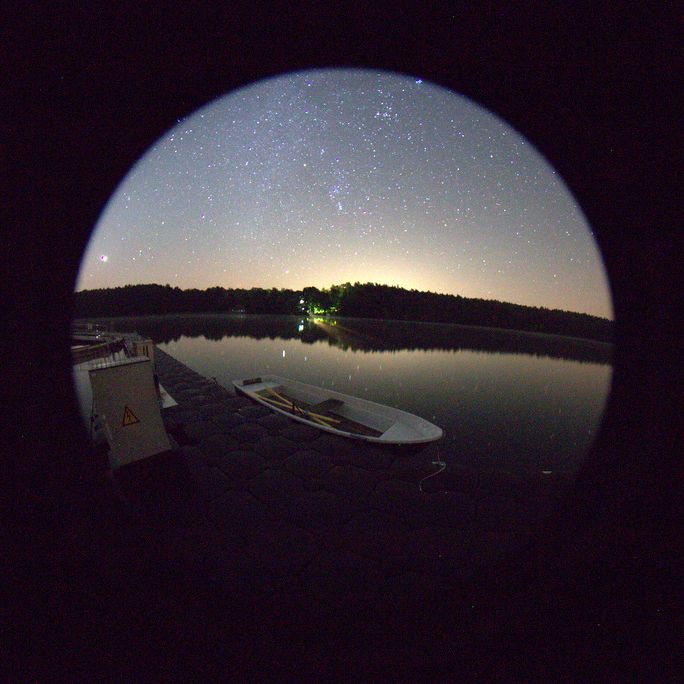}

(c) \includegraphics[width=5.4cm]{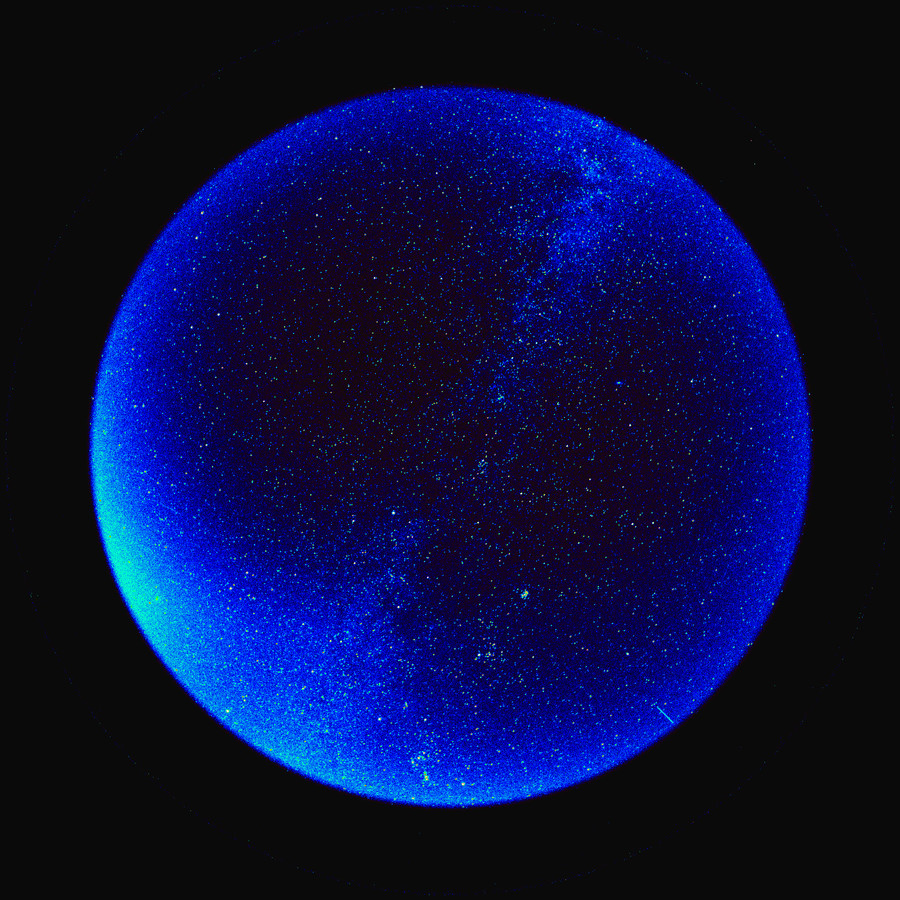}
(d) \includegraphics[width=5.4cm]{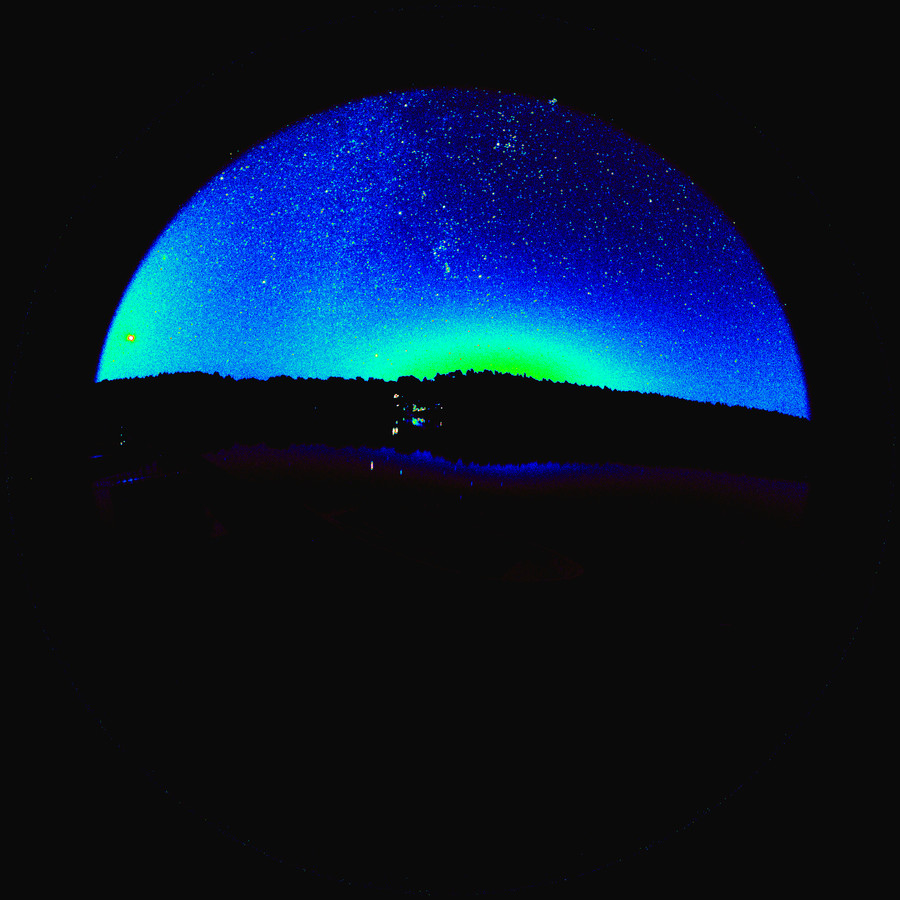}
\includegraphics[width=1.2cm]{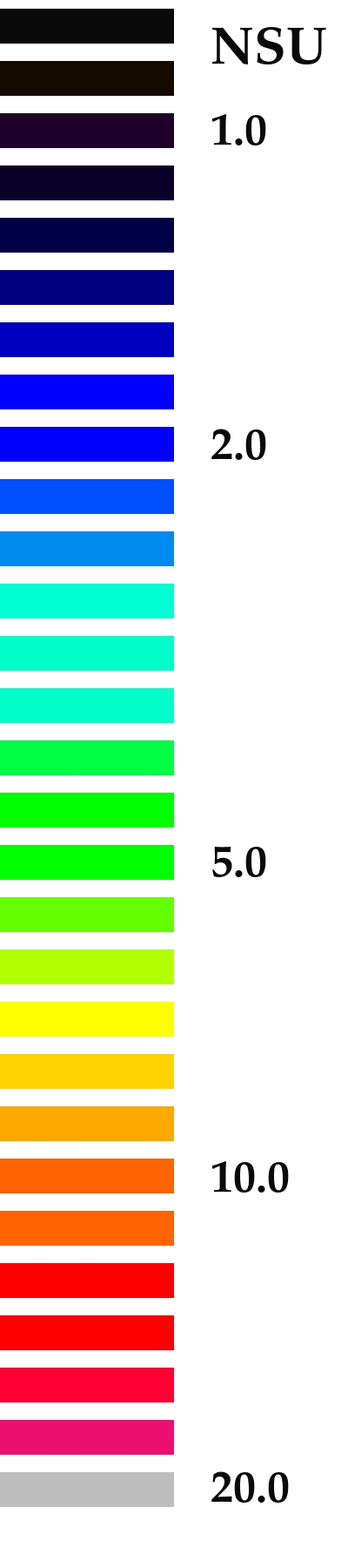}

(e) \includegraphics[width=5.4cm]{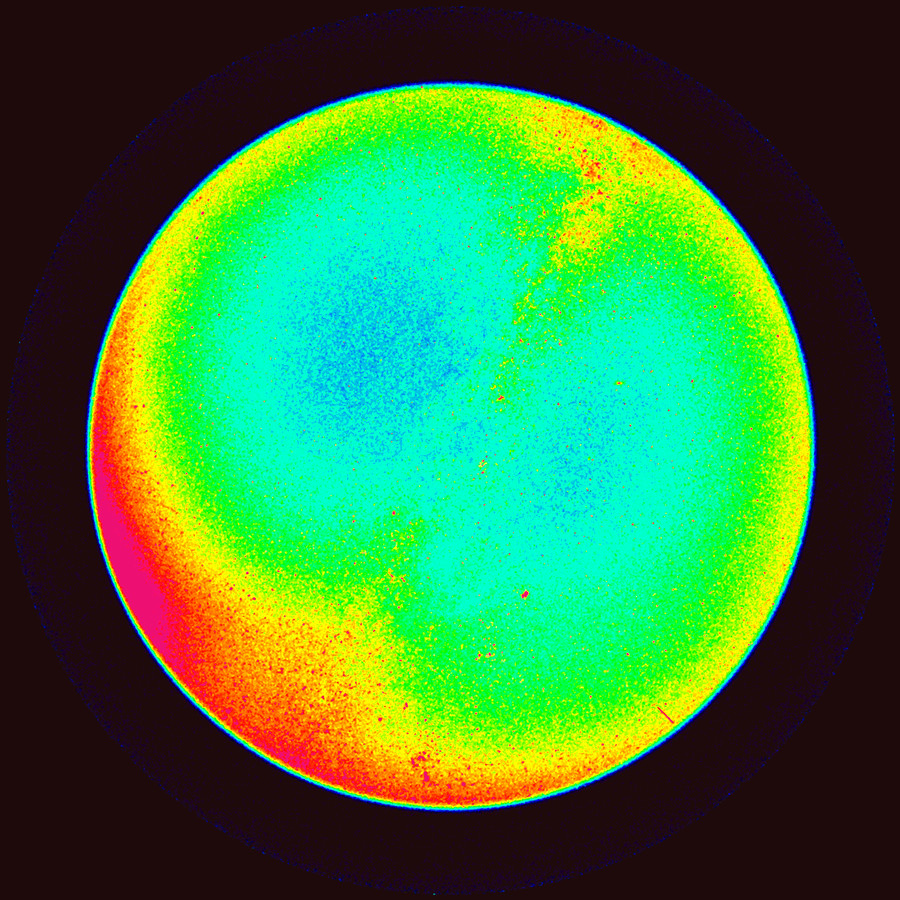}
(f) \includegraphics[width=5.4cm]{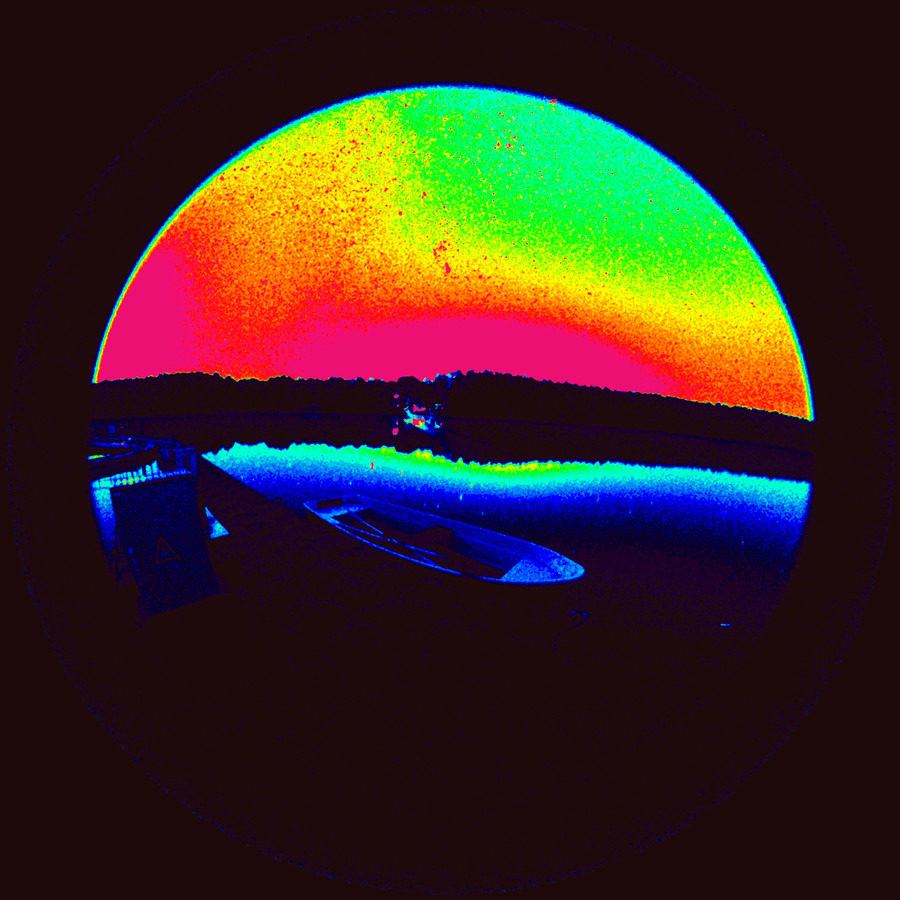}
\includegraphics[width=1.2cm]{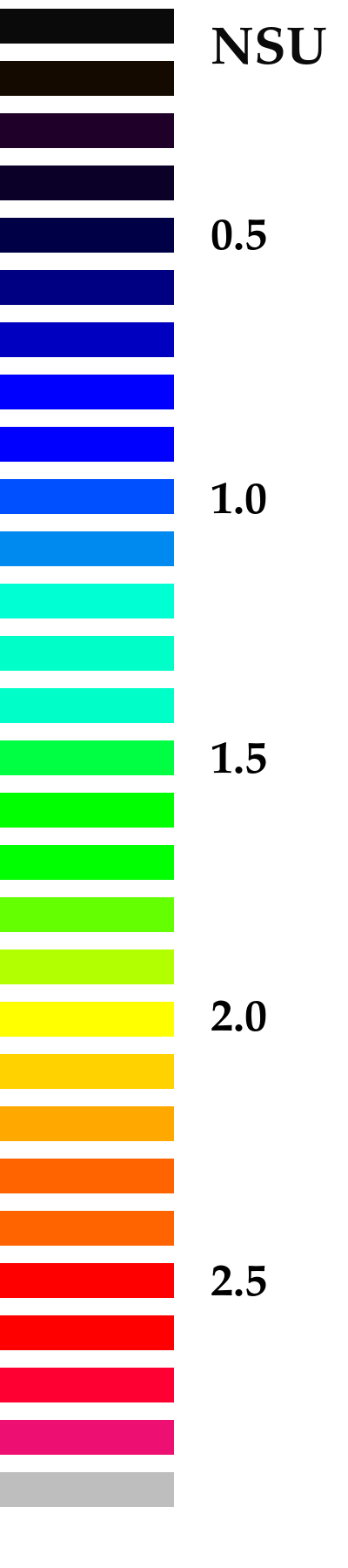}
\caption{All-sky photo of the night sky at Lake Stechlin taken by pointing the camera to the zenith (a), horizontal fish-eye lens photo taken by pointing the camera to the southern horizon (b), and false-color images of these photographs (c-f) showing night sky brightness in natural sky units (NSU).}
\label{allsky1}
\end{figure}

\begin{figure}
(a) \includegraphics[width=5.4cm]{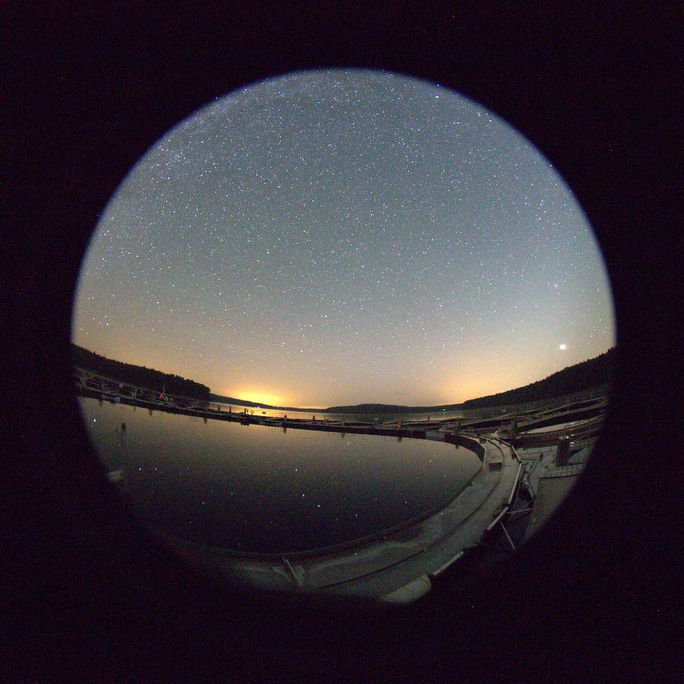}
(b) \includegraphics[width=5.4cm]{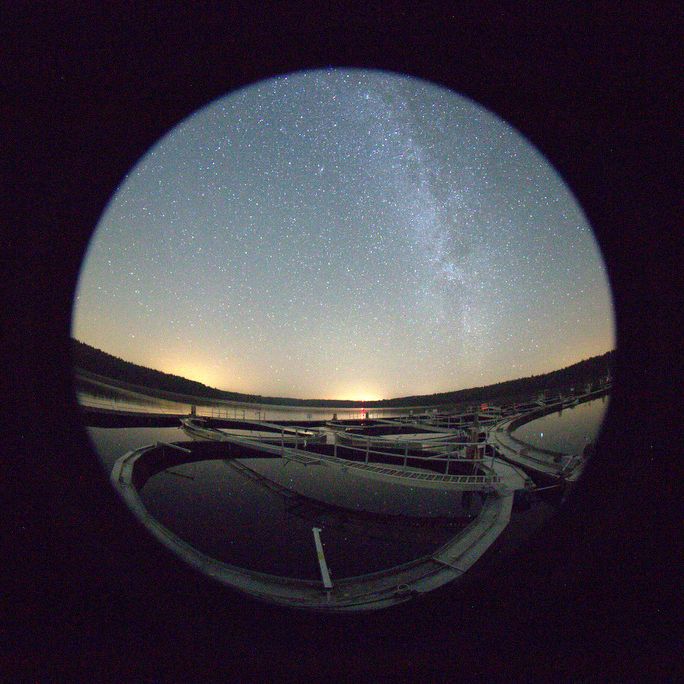}

(c) \includegraphics[width=5.4cm]{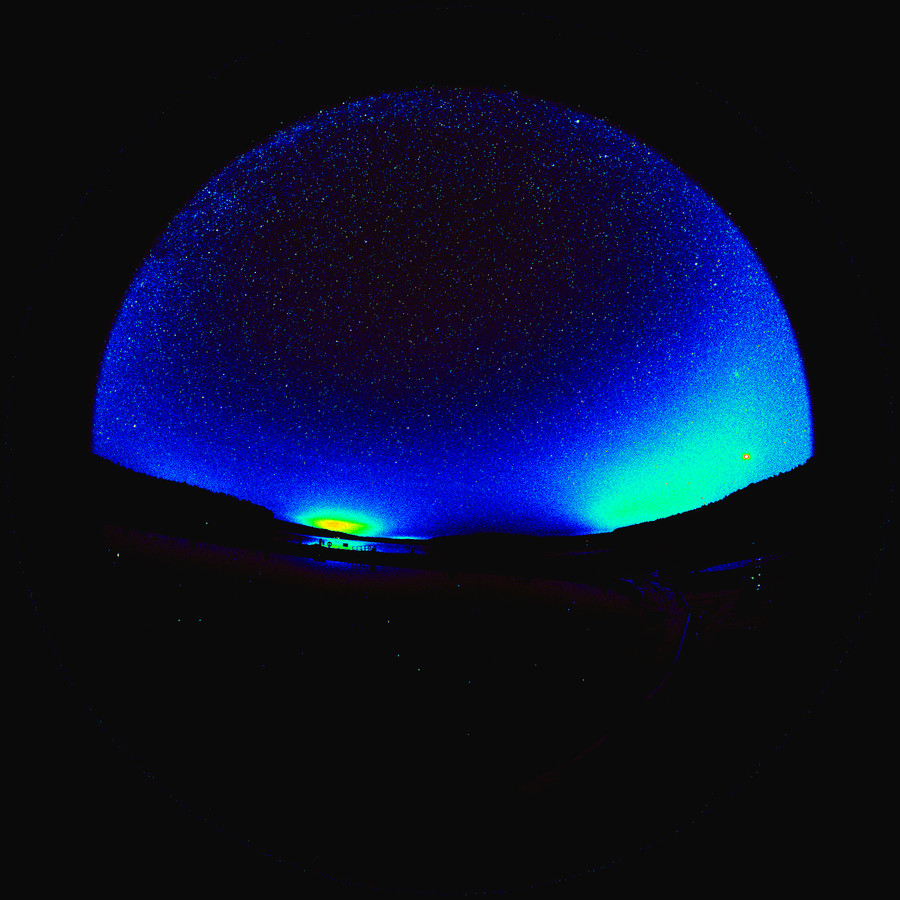}
(d) \includegraphics[width=5.4cm]{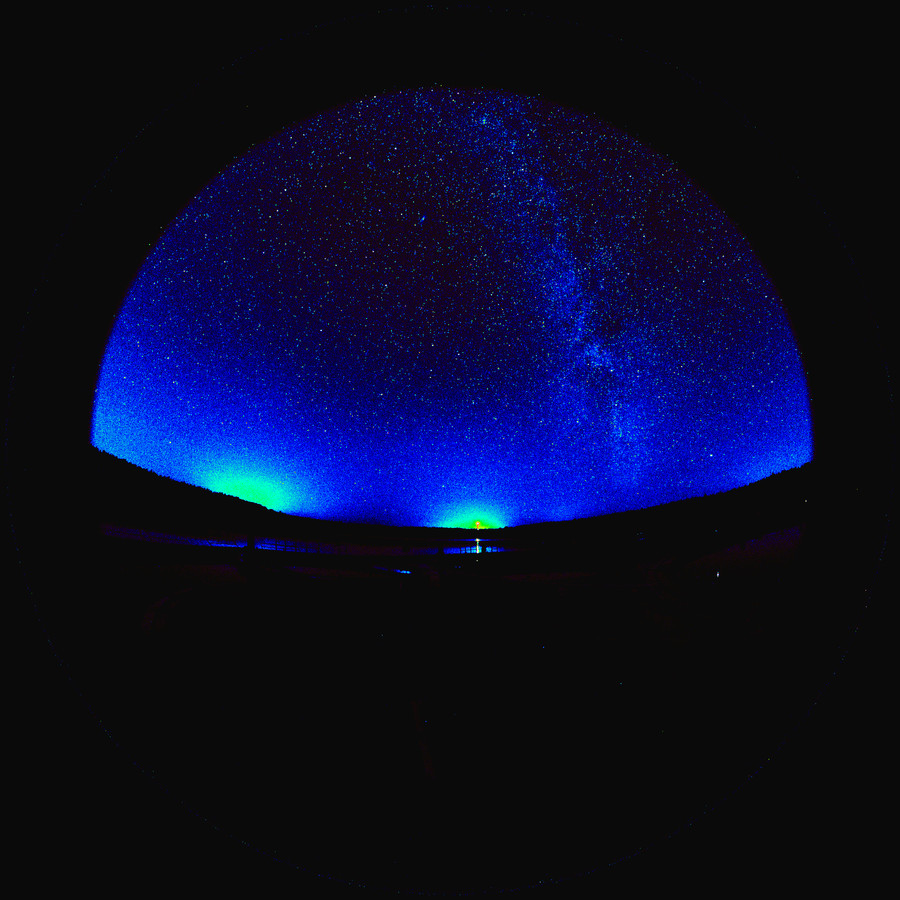}
\includegraphics[width=1.2cm]{colors_log.png}

(e) \includegraphics[width=5.4cm]{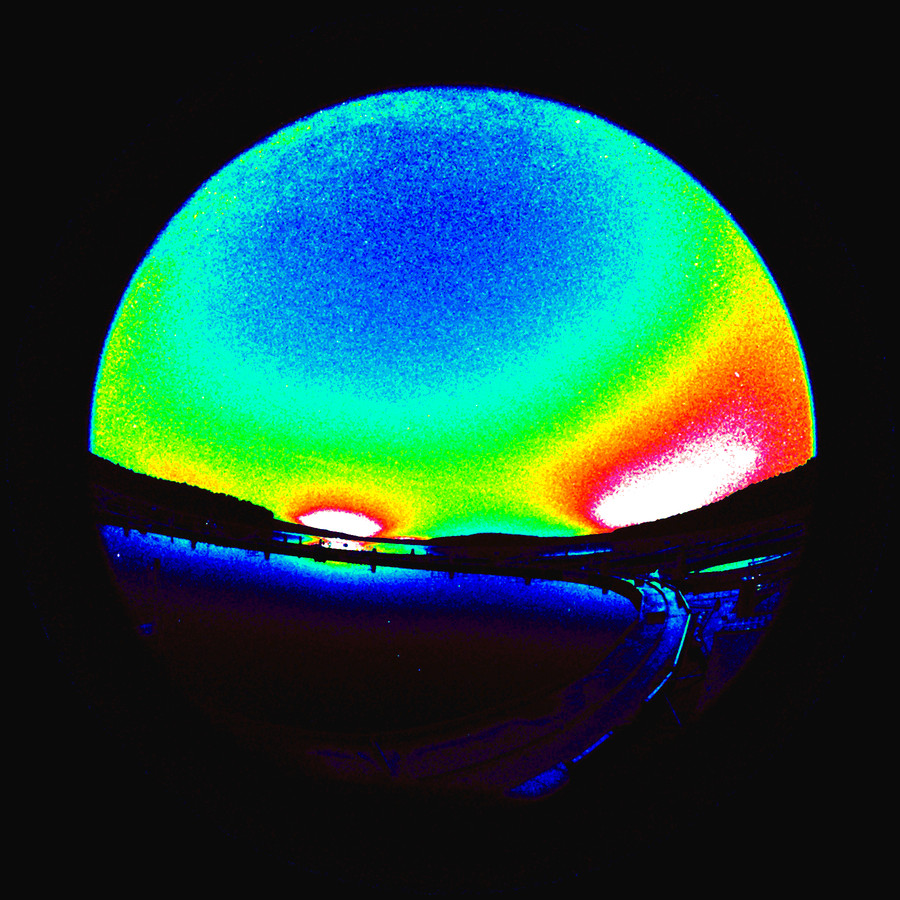}
(f) \includegraphics[width=5.4cm]{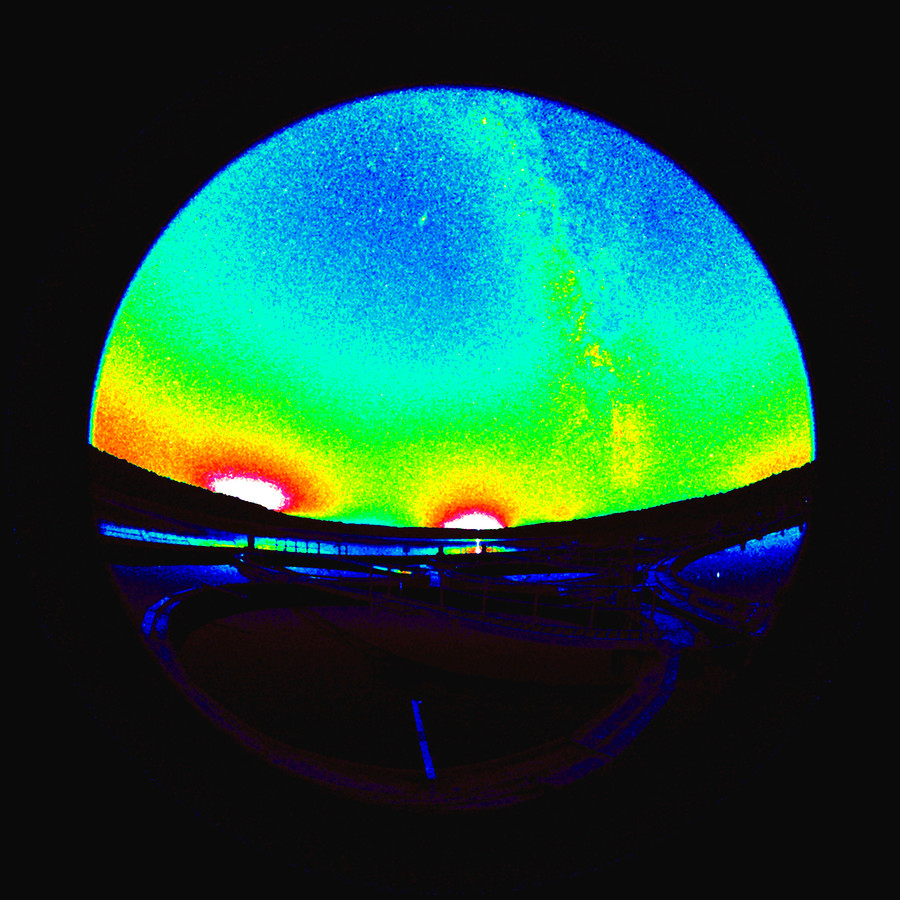}
\includegraphics[width=1.2cm]{colors_lin.png}
\caption{Horizontal fish-eye lens photos of the night sky (a, b) and false color images of these photos (c-f) of the night sky brightness in natural sky units (NSU) showing the view from the LakeLab towards the north (a, c, e) and to the west (b, d, f).}
\label{allsky2}
\end{figure}

The pictures were taken at Lake Stechlin close to the position of the permanent SQM but on a large pontoon (53$^{\circ}$8$^{\prime}$35.2$^{\prime\prime}$N, 13$^{\circ}$1$^{\prime}$41.0$^{\prime\prime}$E) to avoid blurring of the pictures by rocking of the the floating LakeLab construction. The camera was pointed to the zenith to obtain an all-sky image (Fig. \ref{allsky1} a). In addition, the camera was pointed approximately towards the horizon in different azimuth directions to obtain undistorted images from distant light sources near the horizon (Fig. \ref{allsky1} b, Fig. \ref{allsky2}). Several dark frames were taken under the same measurement conditions by leaving the lens carefully covered so that no light could penetrate the optics and reach the detector. Weather conditions allowed data taking at a clear sky and very low amount of wind. No swaying of the floating construction was noted. The Milky Way was overhead and brightly visible by the naked eye.

The original photographs and the calculated false-color images showing the NSB luminance are depicted in Fig. \ref{allsky1} and \ref{allsky2}. In each figure, the upper rows (a and b) show the actual photographs in true color but with enhanced contrast. The middle rows (c and d) show false-color images over the full dynamic range from 0.1 to 20 NSU on a logarithmic scale. The lower rows (e and f) show false-color images on a fine linear scale between 0.1 NSU and 3 NSU.

The left column of Fig. \ref{allsky1} (a, c and e) shows the zenith night sky at 04:57 H. The upper left corner of the pictures points approximately to the north and the lower left corner points eastwards. The structure of the Milky Way is clearly visible in Fig. \ref{allsky1} (a), and Figs. \ref{allsky1} (c) and (e) clearly show that the NSB at the zenith is dominated by the Milky Way. A distinct increase in NSB is apparent on the horizon.

In the right column of Fig. \ref{allsky1} (b, d and f) the horizontal view from the LakeLab towards the south at 4:44 H is shown. In the upper picture (Fig. \ref{allsky1} b), the lit buildings of the IGB on the shore of Lake Stechlin can be seen as well as a glow in the distance. The false color images reveal a bright area in the center of the picture with an NSB of about 4 NSU. This is likely due to the town of Gransee (about 6,000 inhabitants, 53$^{\circ}$0$^{\prime}$N 13$^{\circ}$9$^{\prime}$E, distance ca. 18 km). Furthermore, a larger distant circular region with NSB values of 2 NSU is presumably due to Berlin at a distance of about 70 km. The bright region on the left side is the town of F{\"u}rstenberg (about 6,000 inhabitants, 53$^{\circ}$11$^{\prime}$N 13$^{\circ}$9$^{\prime}$E, distance ca. 9 km).

The left column of Fig. \ref{allsky2} (a, c and e) shows the horizontal view from the LakeLab towards the east at 04:31 H. Two bright orange areas are apparent on the horizon (Fig. \ref{allsky2} a). The area on the left is the town of Neustrelitz (about 20,000 inhabitants, 53$^{\circ}$22$^{\prime}$N 13$^{\circ}$4$^{\prime}$E, distance ca. 25 km) and the area on the right is the town of F{\"u}rstenberg. The area around Neustrelitz had a NSB of up to 7 NSU while the NSB reached about 3 NSU in the direction of F{\"u}rstenberg. The false-color images (Fig. \ref{allsky2} c,e) reveal a less pronounced bright area on the horizon left of Neustrelitz, which is presumably due to the town of Wesenberg (about 3,000 inhabitants, 53$^{\circ}$16$^{\prime}$N 12$^{\circ}$58$^{\prime}$E, distance ca. 15 km) as well as a small bright strip east of Neustrelitz, which we assume to be the town of Neubrandenburg (about 66,000 inhabitants, 53$^{\circ}$33$^{\prime}$N 13$^{\circ}$16$^{\prime}$E, distance ca. 45 km). The sharp bright spot on the right hand side is Sirius. 

In the right column of Fig. \ref{allsky2} (b, d and f) the horizontal view from the LakeLab towards the west at 4:35 H is shown. The red spot in the center of the upper picture (Fig. \ref{allsky2} b) is the tower of the former nuclear power plant of Rheinsberg. The distant orange glow behind it is presumably the town of Wittstock (about 15,000 inhabitants, 53$^{\circ}$10$^{\prime}$N 12$^{\circ}$30$^{\prime}$E, distance ca. 35 km) with a NSB of 4 NSU. The bright spot on the left side is the town of Rheinsberg (about 8,000 inhabitants, 53$^{\circ}$6$^{\prime}$N 12$^{\circ}$53$^{\prime}$E, distance ca. 10 km).

Surprisingly, the closest village, Neuglobsow, is not clearly resolvable in these pictures, which may be attributed to the dense forest vegetation surrounding the village and the absence of industrial areas, shopping centers or similar infrastructure. Furthermore, part-night street lighting is applied in Neuglobsow: only half of the luminaires are lit during the night.

From our data, we evaluate the influence of skyglow originating from Berlin to be low compared to nearby settlements. This is in agreement with a long term NSB study performed in Potsdam (160,000 inhabitants) \cite{Puschnig:2014_Potsdam}, located at 28 km distance from Berlin and where the LP was already reduced compared to the metropolitan center.

\section{Summary and conclusion}
We investigated the NSB at a research field site situated on a lake 70 north of the city of Berlin. Continuous monitoring between the beginning of June and the end of September 2015 was performed using a small photometer (an SQM) collecting data every 10 minutes. In addition, fish-eye lens imaging was performed on a moonless clear night to spatially resolve and quantify distant skyglow caused by LP. 

With the SQM, zenith NSBs as low as 0.32 NSU (22.84 mag$_{SQM}/$arcsec$^2$)  were found under overcast conditions. On clear moonless nights, values of about 1 NSU (21.5-21.7 mag$_{SQM}/$arcsec$^2$) were obtained. The darkening in the presence of clouds indicates that the zenith NSB was not strongly influenced by LP. In areas with LP, a brightening would be observed as shown by simultaneous suburban SQM measurements. Furthermore, the circa-lunar NSB variation was clearly perceptible, which is usually blurred or even wiped out when monitoring zenith NSB in the presence of strong LP.

All-sky and horizontal fish-eye lens imaging revealed several nearby and distant towns as sources of LP measurable on Lake Stechlin. However, the NSB values were below 5 NSU and the brightness was confined to areas close to the horizon.
 
Overall, the data indicates that the influence of skyglow caused by LP is minor at Lake Stechlin during summer and fall, suggesting that this field site is suitable to study the influence of skyglow on lake ecosystems.

\section{Acknowledgements}
This work was supported by the ILES project funded by the Leibniz Association, Germany (SAW-2015-IGB-1), the ``Verlust der Nacht'' project funded by the Federal Ministry of Education and Research, Germany (BMBF-033L038A) and the EU COST Action ES1204 (Loss of the Night Network). The DSLR camera was provided by the GFZ German Research Centre for Geoscience. The authors would like to thank the ILES team for fruitful discussions and Stefan Heller, Armin Penske and Michael Sachtleben for help with the SQMs.


\section{References}



\end{document}